\documentstyle[12pt]{article}
\begin{document}

\begin{titlepage}
\begin{flushright}
CERN-TH.7089/93\\
IOA 297\\
NTUA 42/93\\
\end{flushright}
\vspace{.2in}
\begin{centering}
\vspace{.1in}
{\large {\bf One loop corrections to the Neutralino sector
and Radiative Electroweak Breaking in the MSSM}}\\
\vspace{.4in}
{\bf A.B.Lahanas\hspace{.1cm}}$^{\ddag }$\\
\vspace{.05in}
Theory Division, CERN, CH-1211, Geneva 23, Switzerland.\\
\vspace{.05in}
and \\
\vspace{.05in}
{\bf K.Tamvakis}\\

\vspace{.05in}
Division of Theoretical Physics, Physics Department,\\
University of Ioannina, Ioannina GR-541 10, GREECE \\
\vspace{.05in}
and \\
\vspace{.05in}
{\bf N.D.Tracas}\\
\vspace{.05in}
Physics Department, National Technical University,\\
GR-157 80 Zografou, Athens, GREECE \\
\vspace{.3in}
{\bf Abstract}\\
\vspace{.05in}
\end{centering}
{\small
We compute one loop radiative corrections to the physical
neutralino masses in the MSSM considering the dominant top-stop
contributions. We present a numerical renormalization
group analysis of the feasibility of the radiative gauge
symmetry breaking parametrized by the standard soft
supersymmetry breaking terms. Although the above computed
effects can be in principle large for extreme values of the
Yukawa couplings, they do not in general exceed a few per cent
for most of the parameter space. Therefore tree level constraints
imposed on the gluino mass $m_{\tilde{g}}$
and on the superpotential parameter $\mu$ by LEP1 and CDF
experiments, are not upset by the heavy
top-stop sector.}

\paragraph{}
\par
%
\vspace{.23in}
\begin{flushleft}
CERN-TH.7089/93\\
IOA 297\\
NTU 42/93\\
November 1993
\end{flushleft}
\vspace{.1in}
\rule[.01in]{3in}{.01in}\\
$^{{\hspace{.3cm}}\ddag}${\small \hspace{.1cm}On leave of
absence from\hspace{.2cm}:\hspace{.2cm}Physics Department,
Nuclear and Particle Physics Section,\\University of Athens,
GR-157 71 Athens,\hspace{.2cm}GREECE.}
\end{titlepage}


Supersymmetry seems to be the only framework which allows unification
of the three gauge coupling constants at a common energy scale, while
at the same time respects their low energy values as well as the lower
bounds on proton decay $^{\cite{amal,ekn,lang}}$.
Moreover, softly broken supersymmetry (resulting possibly from an
underlying superstring framework), could lead to $SU(2)_L\times
U(1)_Y$ gauge symmetry breaking through radiative corrections, for
a certain range of values of the parameters
$^{\cite{ir,ikkt,iba,klnq}}$.
Thus, the elegant ideas
of supersymmetry, gauge coupling unification and radiative symmetry
breaking could be realized within the same framework. The Minimal
Supersymmetric Standard Model (MSSM) incorporates all of the above.
It has recently attracted a lot of attention and it has been the subject
of a numerous analyses based on the renormalization\break
group
$^{\cite{rr,klpny,an,op,ram}}$.
It has also recently become evident that, due to the largeness of the
top quark mass, the one loop contributions to the Higgs potential and
to the Higgs physical masses could be important
$^{\cite{erz,oyy,hh,brig,kkw}}$.
Then it is possible that the sparticle masses could also acquire non
negligible radiative corrections from the top-stop contributions.
Since the neutralino sector is, in general, the lightest sector of the
theory (accommodating the LSP), with a possible exception of a Higgs,
it seems to be a good place to search for
substantial radiative corrections.

In the present paper we perform a one loop calculation of the neutralino
physical masses in the context of the MSSM. Only the dominant top-stop
contributions are taken into account. We also perform a renormalization
group analysis of the model to determine the range of parameters that
leads to radiative electroweak breaking with the correct value of
$M_Z$. Our numerical analysis follows that of refs\cite{rr,ram}.
As inputs we consider the parameters $A_0$, $m_0$ and $m_{1/2}$, which
parametrize the soft supersymmetry breaking terms, $\tan \beta (M_Z)=
v_2/v_1$ (the ratio  of the two Higgs field v.e.v.'s) and the running
top quark mass $m_t(M_Z)$ at the scale $M_Z$. The parameters $B(M_Z)$
and $\mu(M_Z)$, which set the size of the mixing of the Higgs scalars
and Higgsinos, can be derived by minimizing the Higgs
potential. The value of the parameter $\mu$, which is essential for
the neutralino masses, is sensitive to radiative effects and thus the
one loop effective potential should be used in the minimization
procedure.

LEP1 and Tevatron CDF experiments put constraints on the parameter space
\break
$(m_{1/2},\mu)$, or equivalently $(m_{\tilde{g}},\mu)$, where
$m_{\tilde{g}}$
is the gluino mass. In these analyses the effects of the radiative
corrections to the neutralino masses and especially to the LSP
$\tilde{Z_1}$, produced e.g. in $Z\rightarrow \tilde{Z_1}\tilde{Z_1},$
have been ignored. The possible existense of a region in the parameter
space where the top-stop radiative corrections become important, means
that the experimental bounds should be reconsidered taking into account
those effects. On the other hand, if these radiative effects are small
the tree level bounds on $m_{1/2}$ and $\mu$ can be trusted
$^{\cite{ua1,bt}}$.

The superpotential of the MSSM is $^{\cite{nil,ln}}$
\begin{equation}
{\cal W}=(h_U\hat{Q}^i\hat{H}_2^j\hat{U}^c
       +h_D\hat{Q}^i\hat{H}_1^j\hat{D}^c
       +f_L\hat{L}^i\hat{H}_1^j\hat{E}^c
        +\mu \hat{H}_1^i\hat{H}_2^j)\epsilon _{ij},
\quad\quad \epsilon_{12}=+1,
\end{equation}
where $\hat{Q}$, $\hat{D}^c$, $\hat{U}^c$, $\hat{L}$, $\hat{E}^c$,
$\hat{H}_1$ and $\hat{H}_2$ stand for the $(3,2,1/6)$,
$(\bar 3,1,1/3)$, $(\bar 3,1,-2/3)$, $(1,2,-1/2)$, $(1,1,1)$,
$(1,2,-1/2)$ and $(1,2,1/2)$ matter chiral superfields. Colour and
family indices are suppressed. The only dimensionful parameter is
$\mu$. Our analysis will be independent of the exact dynamical origin
of this parameter $^{\cite{kn,gm}}$
as long as it has the right order of magnitude (${\cal O}(M_W)$).

The scalar potential involves soft SUSY breaking terms given by
\begin{eqnarray}
V_{sb}&=&m^2_{H_1}|H_1|^2+m^2_{H_2}|H_2|^2\nonumber\\
&+&m^2_Q|\tilde Q|^2
          +m^2_{U^c}|\tilde U^c|^2+m^2_{D^c}|\tilde D^c|^2
      +m^2_L|\tilde L|^2+m^2_{E^c}|\tilde E^c|^2\nonumber\\
      &+& (h_UA_U\tilde QH_2\tilde U^c+h_DA_D\tilde QH_1\tilde D^c
+h_LA_L\tilde LH_1\tilde E^c+h.c.)
           +(B\mu H_1H_2+h.c.).
\end{eqnarray}
We also have the soft breaking Majorana masses for the gauginos
\begin{equation}
{\cal L}_{sb}=-\frac{1}{2}\sum_A M_A{\bar{\lambda}}_A\lambda_A.
\end{equation}

Since radiative corrections are expected to be small, whenever the
large top quark Yukawa coupling is not involved, a reasonable
approximation is to keep only loops which involve the top-stop
system. In that case the one loop corrections to the scalar potential
are
\begin{equation}
\Delta V_1=\frac{3}{64\pi^2}
    \sum_{+,-}m^4_{\pm}\left[\ln\frac{m^2_{\pm}}{Q^2}-\frac{3}{2}\right]
-\frac{3}{32\pi^2}m^4_t\left[\ln\frac{m^2_t}{Q^2}-\frac{3}{2}\right],
\end{equation}
where the $\overline{DR}$ regularization scheme has been used. Taking
all fields to vanish, except the neutral Higgses, we have
\begin{equation}
m_t\equiv h_tH^0_2,\quad\quad
m^2_\pm\equiv\frac{1}{2}\left[m^2_{LL}+m^2_{RR}\pm
    \left(\left(m^2_{LL}-m^2_{RR}\right)^2
                    +4m^4_{LR}\right)^{1/2}\right],
\end{equation}
where
\begin{eqnarray}
m^2_{LL}&\equiv & m^2_{\tilde t}+m^2_t+
                 \left(\frac{g^{\prime ^2}}{12}-\frac{g^2}{4}\right)
                 \left(|H^0_2|^2-|H^0_1|^2\right),\nonumber\\
m^2_{RR}&\equiv & m^2_{\tilde t ^c}+m^2_t-
                 \frac{g^{\prime ^2}}{3}
                 \left(|H^0_2|^2-|H^0_1|^2\right),\nonumber\\
m^2_{LR}&\equiv & h_t\left(H^0_2A+\mu H^{0^*}_1\right).
\end{eqnarray}
All parameters are considered to be running ones, depending on the
scale Q appearing in eq.(4)

Minimization of the scalar potential yields two conditions on the
Higgs v.e.v.'s \break
$v_2\equiv <H_2>$ and $v_1\equiv <H_1>$
\footnote{We follow the conventions of J. Ellis and F. Zwirner, Nucl.
Phys. B338(1990)317, where $B$ and $\mu$ have opposite signs.}
\begin{eqnarray}
\frac{M_Z^2}{2}&=&\frac{{\bar m}_1^2-{\bar m}_2^2 \tan ^2\beta}
                        {\left(\tan ^2\beta-1\right)},
\quad\quad \left(M_Z=91.2{\rm GeV}\right)\\
\sin 2\beta &=& -\frac{2B\mu}{{\bar m}^2_1+{\bar m}^2_2}\quad .
\end{eqnarray}
In eqs.(7,8) we have defined $^{\cite{ram,an1}}$
\begin{equation}
{\bar m}^2_{1,2}\equiv m^2_{1,2}+\frac{\partial \Delta V}
                                      {\partial v^2_{1,2}},
\quad\quad {\rm with}\quad m^2_{1,2}\equiv m^2_{H_{1,2}}+\mu ^2.
\end{equation}

The key features of the radiative symmetry breaking could be seen even
with the tree level potential. This breaking occurs at the scale
$Q_b$ where the ($Q$-dependent) expression $m^2_1m^2_2-(\mu B)^2$
becomes negative. The free parameters of the model should be chosen in
a manner consistent with the observed value of $M_Z$. At another scale
$Q_c<Q_b$ the expression $m^2_1+m^2_2-2\mu B$ becomes negative.
This makes the tree level potential unbounded from below and therefore
untrustworthy. The above consideration makes the choice of the scale
$Q$ critical. In contrast, the inclusion of the one loop correction to the
potential makes this choice irrelevant, as long as it is near
$M_Z$. The reason is that the one loop corrected potential is
to this order of computation, renormalization (i.e. $Q$) independent,
up to constant (but $Q$-dependent) terms which do not appear in
the minimization equations. Therefore, using the one loop corrected
potential, we consider the minimization conditions, eqs(7,8) at
$Q=M_Z$. Given a set of values for the input parameters at some scale
$M_{GUT}$, the validity of eqs(7,8) ensures that the electroweak
breaking occurs with the correct value for $M_Z$. As we have already
mentioned above, we take $\tan \beta (M_Z)$ and $m_t(M_Z)$ as inputs.
In that case the parameters $B(M_Z)$ and $\mu (M_Z)$ are derived from
the eqs(7,8). The connection between the values of the parameters at
the scale $M_{GUT}$ and those at $M_Z$ is carried out by the
renormalization group equations. More details on the numerical
procedure followed will be given below.

The neutral gauge-fermion and neutral Higgsino (collectively
``neutralino'') mass-matrix is easily seen to be
\begin{equation}
{\cal M}=\left(\begin{array}{cccc}
M_1      &     0     & g^\prime v_1/\sqrt 2   &  -g^\prime v_2/\sqrt 2\\
 0        &    M_2    & -g v_1/\sqrt 2          &  g v_2/\sqrt 2\\
g^\prime v_1/\sqrt 2&
            -g v_1/\sqrt 2&     0  &  -\mu\\
-g^\prime v_2/\sqrt 2&
                      g v_2/\sqrt 2&  -\mu &  0
\end{array}\right)
\end{equation}
where we have used a $\psi^i\equiv\left(\tilde B,\tilde W_3,\tilde
\chi _1,\tilde \chi _2\right)$ Weyl basis, with $\tilde \chi
_{1,2}\equiv i\tilde H_{1,2}$ to make ${\cal M}$ real and symmetric.
Radiative corrections to the neutralino two point function take the
form $-i(p\hspace{-2mm}/^{\alpha\.{\beta}})a_{ij}$ and
$-i\epsilon^{\beta\alpha}b_{ij}$
($p\hspace{-2mm}/^{\alpha\.{\beta}}
\equiv p^{\mu}
\sigma_{\mu}^{\alpha\.{\beta}}$). Taking into account only the dominant
top-stop contributions (Fig.1) we obtain the following expressions
\begin{eqnarray}
a_{\tilde B\tilde W}&=&
  \frac{1}{2}\sqrt{\frac{3}{5}}\frac{(\alpha_1\alpha_2)^{1/2}}{4\pi}
           \left(\Lambda_+\cos ^2\phi+\Lambda_-\sin ^2\phi\right)
\nonumber\\
&=&\left(\frac{\alpha_1}{15\alpha_2}\right)^{1/2} a_{\tilde W\tilde W}
\nonumber\\
a_{\tilde B\tilde \chi _2}&=&
  -3\sqrt{\frac{3}{10}}\frac{(\alpha_1\alpha_t)^{1/2}}{4\pi}
                  \cos \phi\sin \phi (\Lambda_+-\Lambda_-)
\nonumber\\
&=&-\sqrt{\frac{3\alpha_1}{5\alpha_2}}a_{\tilde W\tilde \chi _2}
\nonumber\\
a_{\tilde\chi _2\tilde\chi _2}&=&
     \frac{3}{2} \frac{\alpha_t}{4\pi}(\Lambda_++\Lambda_-)
\nonumber\\
a_{\tilde B\tilde B}&=&
\frac{8}{15}\left(\frac{\alpha_1}{\alpha_t}\right)a_{\tilde\chi _2\tilde\chi
_2}
-\left(\frac{15\alpha_1}{\alpha_2}\right)^{1/2}a_{\tilde B\tilde W}
\end{eqnarray}
and
\begin{eqnarray}
b_{\tilde B\tilde B}&=&
\frac{4}{5}\frac{\alpha_1}{4\pi} m_t\cos \phi \sin \phi(L_+-L_-)
\nonumber\\
&=&-\frac{2}{15}\frac{\alpha_1}{\alpha_t}
b_{\tilde\chi _2\tilde\chi_2}=
\frac{2}{\sqrt{15}}\left(\frac{\alpha_1}{\alpha_2}\right)^{1/2}
                                b_{\tilde B\tilde W}
\nonumber\\
b_{\tilde B\tilde\chi _2}&=&
\sqrt{\frac{3}{10}}\frac{(\alpha _1\alpha _t)^{1/2}}{4\pi}m_t
\left[4(L_++L_-)-5(L_+\cos ^2\phi +L_-\sin ^2\phi )\right]
\nonumber\\
b_{\tilde W\tilde \chi _2}&=&
-\frac{3}{\sqrt 2}\frac{(\alpha _2\alpha _t)^{1/2}}{4\pi}m_t
  \left[L_+\cos ^2\phi +L_-\sin ^2\phi \right]
\end{eqnarray}
Note that $a_{ij}=a_{ji}$ and $b_{ij}=b_{ji}$.
Entries which are not shown vanish. In the above formulae
$\alpha_{1,2}$ are the gauge couplings, $\alpha_t\equiv h_t^2/(4\pi)$,
$\phi$ is the angle diagonalizing the stop matrix given by
\[
\cos ^2 \phi=
\frac{m^2_+-m^2_{{\tilde t}^c}-m^2_t-\frac{2}{3}\sin ^2\theta_W M^2_Z
\cos 2\beta}{m^2_+-m^2_-},
\]
and
\begin{eqnarray*}
\Lambda_{\pm}&=&{\rm Re}\int_0^1 dx
   x\ln\left[\frac{p^2}{Q^2}x^2+\frac{m^2_{\pm}-m^2_t-p^2}{Q^2}x
                    +\frac{m^2_t}{Q^2}\right],\\
L_{\pm}&=&{\rm Re}\int_0^1 dx
   \ln\left[\frac{p^2}{Q^2}x^2+\frac{m^2_{\pm}-m^2_t-p^2}{Q^2}x
                    +\frac{m^2_t}{Q^2}\right].
\end{eqnarray*}
Finally $m_{\pm}$ are the physical stop masses and $m_t$ is the running
top mass. They can be read off from eqs(5,6) by putting $H_1$ and
$H_2$ on their v.e.v.'s. We have evaluated them at the scale $Q$
appearing in the integrals $\Lambda$ and $L$. The choice of $Q$ is not
critical, since physical masses, that is poles of the propagators,
should be scale-independent, up to this order of approximation. In our
numerical analysis we have chosen this scale to be the heaviest of the
thresholds involved which is either the gluino mass $m_{\tilde g}$ or
the mass of the  ${\tilde d}^c_{L}$  squark, depending on the values of
the inputs $m_{1/2}$ and $m_0$. We have numerically checked that our
results remain independent of the scale, as far as it lies between
$M_Z$ and the heaviest threshold.

The chirality flipping part of the one loop propagator is found to be
\[
i\epsilon_{\alpha\beta}
\left[\left({\cal M}+b+a{\cal M}+{\cal M}a\right)
\left(p^2-{\cal M}^2-D(p^2)\right)^{-1}\right]_{ij}\quad ,
\]
with
\[
D(p^2)={\cal M}b+b{\cal M}+{\cal M}a{\cal M}+{\cal M}^2a.
\]
Then, physical masses are determined by the poles of the propagator or
equivalently
\begin{equation}
det\left(p^2-{\cal M}^2-D(p^2)\right)=0.
\end{equation}
This is also the position of the poles of the chirality conserving
part. Eq(13) can be solved perturbatively giving corrections $\delta
m^2_n$ to the $n^{\rm th}$ neutralino mass squared $m^2_n$
\begin{equation}
\delta m^2_n=
2\sum_{i,j}R_{ni}R_{nj}\left[m^2_na_{ij}+m_nb_{ij}\right].
\end{equation}
In eq(14) the matrix ${\bf R}$ diagonalizes the tree level neutralino
mass matrix, ${\bf R}{\cal M}{\bf R}^{\top}={\rm diagonal}$ while
$a_{ij}$ and $b_{ij}$ are evaluated for $p^2=m^2_n$. Although it
suffices to know the parameters entering ${\cal M}$ and the stop mass
matrix at the scale $Q$ for the determination of the corrections given
by eq(14), a systematic renormalization group analysis that is
consistent with the radiative breaking scenario and takes into account
all existing experimental and theoretical constraints has to be
performed.

The corrections under discussion, if one neglects the aforementioned
constraints, can be as large as 10\% or even more in some special
cases. In order to understand qualitatively how this may arise,
consider a simplified picture in which the tree level stop mass matrix
is given by
\begin{equation}
{\cal M}^2_{\tilde t}=
\left(\begin{array}{cc}
m^2_L  &  \Delta \\
\Delta &  m^2_R
\end{array}\right)
\end{equation}
with
$m^2_L\simeq m^2_R\simeq m^2_t+m^2_{SUSY}$
and $\Delta =m_t(A+\mu\cot \beta)$. The dominant mass renormalization,
as can be seen from eq(12), is provided by the
$\tilde\chi_2\tilde\chi_2$
term. Obviously sizezable corrections can be obtained if
$b_{\tilde\chi_2\tilde\chi_2}$ becomes large. The latter is true if we have
a large top mass, large mixing ($\cos \phi\sim\sin \phi=1/\sqrt
2$) and large stop mass spliting
(giving $(L_+-L_-)\sim \ln (m^2_+/m^2_-)\sim {\cal O}(1)$). The second
condition is always satisfied since $m^2_L\sim m^2_R$. The effect of
the radiative corrections is expected to be enhanced in the case of
light neutralinos ($<M_W$). If we assume $\tan\beta=1$ one of the
eigenstates has mass $-\mu$. Thus with $|\mu|<M_W$ we always have a
light ($<M_W$) neutralino state. If in addition $M_{1,2}>>M_W$, the
condition $Tr {\cal M}=M_1+M_2$ guarantees that a second light
state exists as well. Thus, in this case we obtain two light
neutralinos. In Table I we present a typical example of this situation
with the resulting radiative corrections to the neutralino tree level
masses, as they are obtained using eq(14)\footnote
{We are aware of the fact that such large values of $m_t$ are
not compatible with $\tan\beta =1$ within the perturbatively regime.},
where all wave function (eq(11)) and mass (eq(12)) renormalization
effects have been taken into account.

One observes a substantial correction (10\%) to the lightest
neutralino state. Therefore, large radiative effects to the light
neutralinos cannot be excluded a priori.
Admittedly this is an oversimplified picture. We know that $m^2_L\neq
m^2_R$ since soft masses $m_{\tilde t}$ and $m_{{\tilde t}^c}$
are different due to their different dependence on
\[
\Delta _t=\int_0^{t_0} dt \frac{h^2_t}{(4\pi)^2}
\left( A^2+m^2_{Q_3}+m^2_{U_3}+m^2_{H_2}\right),
\quad\quad t_0\equiv \ln \frac{m^2_{GUT}}{Q^2}
\]
$m^2_L\sim m^2_R$ can only occur provided that
$m^2_0+7m^2_{1/2}>>2\Delta _t$, that is when the $m_0$ and $m_{1/2}$
dependence of $m^2_{\tilde t,\tilde t ^c}$ overwhelms that of
$\Delta _t$. This constraints restrict considerably the parameter
space. Besides, the large mixing condition we have assumed imposes a
further constraint, namely $m^2_L-m^2_R<<2m_t(A+\mu \cot\beta)$,
reducing even more the allowed region. Even if there is a window where
these conditions are satisfied, consistency with radiative breaking
and small value of $\mu$ (where the radiative effects are important)
is not certain. There is a strong correlation of the parameters
entering in the radiative corrections under discussion and therefore
they cannot be chosen at will. In order to see whether such a
picture can really emerge, we have therefore to perform a systematic
renormalization group analysis that takes into account all theoretical
constraints and experimental bounds put on physical parameters.

The renormalization group equations (RGE) of the running parameters
involved can be found in the literature and will not be repeated
here $^{\cite{ikkt,nil}}$. As free parameters of the model we take the
soft breaking parameters $A_0$, $m_0$, $m_{1/2}$ at the unification
scale $M_{GUT}(\simeq 10^{16}{\rm GeV})$
\footnote {Our numerical analysis reveals that the corrections are not
sensitive to the precise value of $M_{GUT}$ as long as it is in the
range of $10^{16}$GeV. In the rest of our discussion we take
$M_{GUT}= 10^{16}{\rm GeV}$.}
and the values of $\tan\beta (M_Z)$ and $m_t(M_Z)$ as we already
mentioned. This is the same procedure adopted by other authors too.
The value of the Yukawa coupling $h_t(M_Z)$ can easily be evaluated
and the numerical routines for the gauge and Yukawa couplings provide the
value $h_0=h_t(M_{GUT})$ at the unification scale $M_{GUT}$. We shall
limit ourselves to the case where the bottom and the tau Yukawa
couplings are small compared to $h_t$. This, of course, excludes large
values for $\tan\beta$ (some models seem to prefer values in the range
$3<\tan\beta <15$ though higher values $\leq 85$ cannot be excluded)
$^{\cite{na}}$. In any case, our numerical results show that
accepting
nonvanishing bottom and tau Yukawa couplings produces minor changes
in the radiatively corrected neutralino masses, and therefore we
ignore them in the following.

Starting with $A_0$, $m_0$, $m_{1/2}$ and $h_0$ at $M_{GUT}$, we run
the RGE for all soft masses and $A$ down to the scale $M_Z$. The
values $B_0=B(M_{GUT})$ and $\mu _0=\mu(M_{GUT})$ are not considered
as free parameters, as we have repeatedly remarked, but their values
at $M_Z$ can be extracted from the minimization conditions, eq(7,8).
Their values at any other scale can be found by running the RGE for
these parameters. Solving eq(7,8) is a straightforward task, if the
radiative corrections to the potential are ignored. However it is
important that these corrections should be included. In that case the
determination of $B(M_Z)$ and $\mu(M_Z)$ is not that easy
since the dominant top-stop contributions depend on $\mu$ through the
field dependent masses in eq(5,6). Several runs are required to
achieve convergence and get the desired corrected values. It is well
known that the values of $\mu$ obtained by that way could differ
substantially from their tree level ones in some regions of the
parameter space. As far as the boundary conditions at the unification
scale $M_{GUT}$ are concerned, various types are possible depending on
the underlying theory. The simplest choice at $M_{GUT}$, corresponding
to unification within the minimal supergravity context, is
\[
m^2_Q=m^2_{U^c}=m^2_{D^c}=m^2_{E^c}=m^2_L=m^2_{H_1}=m^2_{H_2}=m^2_0.
\]
and
\[M_1=M_2=M_3=m_{1/2},
\]
which we shall assume.

Due to an infrared fixed point $^{\cite{pr,z}}$ of the Yukawa coupling,
$\tan\beta (M_Z)$ is forced to a minimum value, given the input
value $m_t(M_Z)$. If this value is exceeded, $h_t$ becomes
nonperturbatively large as we increase the scale. We shall therefore
impose the perturbative requirement $h_t^2/(4\pi)\leq{\cal O}(1)$ at
all scales. The running mass $m_t$ and the physical (pole of the
propagator) top quark mass $M_t$ are related by $^{\cite{gbgs}}$
\[
M_t=\frac{m_t(M_t)}{\left(1+\frac{4\alpha _3(M_t)}{3\pi}+...\right)}
\quad .
\]
This takes into account the QCD corrections to the top quark
propagator. In our numerical studies we have taken $M_t>110$GeV which
is the lower experimental bound put on $M_t$.

The choice of the scale $Q$ involved in eq(11,12) through the
integrals $\Lambda$ and $L$, is not important as physical masses
should not depend on it. We have numerically checked that this is the
case. In our calculation we have chosen $Q$ to be the largest of the
thresholds encountered as we run from $M_{GUT}$ down to $M_Z$.
Following ref\cite{rr,ram}, we have ignored the small electroweak breaking
effects and define the threshold $\mu _i$, for the particle $i$, to be
the scale where
\[
m^2_i(\mu ^2_i)=\mu ^2_i .
\]
The scale $Q$ turns out to be the heaviest of the gluino and the
${\tilde d}^c_{L}$ squark mass, depending on the initial
values of $m_0$ and $m_{1/2}$. The electroweak breaking effects do not
alter substantially the value of $Q$, provided it stays
in the TeV range.

In Tables II - V we present sample results for
$m_t(M_Z)=150$ and 180GeV and for various values of $m_0$ and
$m_{1/2}$ such that $m_0\simeq m_{1/2}$, $m_0>m_{1/2}$ and
$m_0<m_{1/2}$. Cases where $A_0=0$ or $A_0\neq 0$ are shown. As
far as the values of $\tan\beta (M_Z)$ are concerned we present results
for the smallest allowed value and for $\tan\beta (M_z)=10$. These
cases give a clear general picture. We have actually
scanned a large region of the parameter space confirming the above
statement. Therefore these results should be considered as
representatives of the general situation. Deliberately and in order to
explore a large region of the parameter space, we have not imposed any
constraints on the trilinear coupling $A$ resulting from the tree
level study of the scalar potential. These constraints, which are
imposed to avoid charge and color breaking minima, are known to be too
restrictive and the picture could very well change when loop effects are
taken into account $^{\cite{gamb}}$. We also do not exclude cases
where the light Higgs gets lower than the ALEPH bound of $\simeq
58$GeV $^{\cite{aleph}}$ or chargino masses lower than 45GeV.

As one can observe, the correction to the neutralino masses are quite
small. They are less than 1\% even in cases where  light states
$\simeq$40GeV appear in the spectrum. In the case of a very light
$\simeq$20GeV neutralino, the corrections are slightly enhanced to
2.5\%. The smallnes of these corrections is not a result of a mere
destructive interference of wave function and mass renormalization
effects. We can switch off the wave function renormalization contribution
to the neutralino propagator and no enhancement is observed
\footnote{ The wave function and mass renormalization effects are
actually found to be of the same order of magnitude.}.
Such a case is shown in Table III. We have also allowed for a
scale $Q$ different from the heaviest threshold. No significant
change, up to this order of approximation, was observed. The smallness
of the radiative corrections under discussion leaves no hope that the
inclusion of the remaining particle contributions to the effective
potential will alter the situation. Two simple arguments are in favour
of this statement: (i) it is known that such contributions could not be
larger than those of the top-stop system and (ii) the scale $Q$ could
not play any critical r\^{o}le since our results do not depend on that
scale.

Let us summarize by repeating our main conclusions.
We have systematically computed the top-stop one loop corrections to
the physical neutralino masses in the MSSM. The heaviness of the top
quark ensures that these corrections are the dominant ones. Although
large corrections, of the order of 10\% cannot be excluded, these
correction turn to be very small if the radiative electroweak breaking
scenario is adopted. This scenario, along with the universal boundary
conditions imposed at $M_{GUT}$ (making the number of free parameters
small), strongly correlates the running parameters and diminishes the
effects of the radiative corrections. In most of the parameter space
these corrections never exceed a few per cent $<{\cal O}(2\%-3\%)$,
even in cases where some of the neutralino masses  are low ${\cal
O}(20{\rm GeV})$ and the effect would have been expected to be
enhanced. This implies that the relevant parameter space for the
phenomenological study of the neutralino sector is consistently
described by the $(m_{\tilde g},\mu)$ pair, since these two quantities
are very weakly correlated to the other parameters of the model and
especially to the heavy top quark mass. This natural suppression of
the radiative effects in the neutralino sector from the top-stop
system, may be wellcome since it shows that LEP and CDF
phenomenological studies and constraints imposed on  $m_{\tilde g}$ and
$\mu$ are stable against radiative corrections.

\vspace{1cm}
Aknowledgements

We would like to thank the CERN Theory Division for its hospitality. Two
of us (A.B.L. and K.T.) would like to thank R. Arnowitt for valuable
discussions. A.B.L. wishes also to thank C. Kounnas and F. Zwirner for
discussions. The work of A.B.L. was partly supported by the EEC
Science Program SC1-CT92-0792. K.T. aknowledges travelling support
by the EEC Contract SCI-0394-L and the University of Ioannina Research
Committee. The work of N.D.T. was partly supported by the CEC Science
Program SCI-CT91-0729.
\vfill\newpage

\newpage

\newpage
%
\begin{center}
{\bf Table I}\\
\vspace{.15cm}

\begin{tabular}{|cccc|}
\hline
$m_t=170$GeV   & $A_0=300$GeV      & $m_0=250$GeV &   $m_{1/2}=800$GeV
$\quad$\\
              & $\tan \beta =1$ &  $\mu =700$GeV     &               \\
\hline
\multicolumn{4}{|c|}{Stop masses= 393GeV, 169GeV}\\
\hline
\multicolumn{2}{|c}{Neutralino masses:}&  &  \\
$m_i$       & $m_i^\prime$     &  $m_i-m_i^\prime $   &  (\%)  \\
670.70      & 669.10           &1.60                 & .24    \\
350.82      & 350.54           & 0.28                & .08    \\
70.00       &  71.98           &-1.98                &-2.83    \\
48.47       &  43.41           &5.06                 &10.44   \\
\hline
\end{tabular}

\vspace{.8cm}
{\bf Table II}\\
\vspace{.15cm}

\begin{tabular}{|cc|cc|}
\hline
\multicolumn{4}{|c|}
{$m_0=200$GeV, $m_{1/2}=100$GeV, $A_0=0(200)$GeV}\\
\multicolumn{4}{|c|}{$m_t(M_Z)=150$GeV, $\tan\beta (M_Z)=10$}\\
\hline
\multicolumn{4}{|c|}{abs. value of $\mu =96(150)$GeV}\\
\hline
\multicolumn{2}{|c}{$\mu>0$} & \multicolumn{2}{|c|}{$\mu<0$}\\
Tree level mass & Correction (\%) & Tree level mass & Correction (\%)\\
\hline
162.67(194.88)  & $+0.06(-0.45)$    & 153.33(184.77)  & $-0.04(-0.44)$\\
115.73(164.78)  & $+0.45(+0.22)$    & 123.10(171.07)  & $+0.37(+0.17)$\\
{}~56.30(~62.71)  & $+0.04(-0.16)$    & ~59.56(~70.97)  & $-0.20(-0.40)$\\
{}~23.00(~33.97)  & $-1.60(-1.00)$    & ~37.02(~42.10)  & $-0.68(-0.19)$\\
\hline
\end{tabular}

\vspace{.8cm}
{\bf Table III}\\
\vspace{.15cm}

\begin{tabular}{|cc|cc|}
\hline
\multicolumn{4}{|c|}
{$m_0=0$GeV, $m_{1/2}=200$GeV, $A_0=0$GeV}\\
\multicolumn{4}{|c|}{$m_t(M_Z)=150$GeV, $\tan\beta (M_Z)=10$}\\
\hline
\multicolumn{4}{|c|}{abs. value of $\mu =246$GeV}\\
\hline
\multicolumn{2}{|c}{$\mu>0$} & \multicolumn{2}{|c|}{$\mu<0$}\\
Tree level mass & Correction (\%) & Tree level mass & Correction (\%)\\
\hline
285.29   & $+0.39(-0.11)$    & 151.81   & $-0.49(-0.26)$\\
254.40   & $-0.29(-0.40)$    & 258.38   & $-0.30(-0.38)$\\
142.18   & $-0.52(-0.27)$    & 275.30   & $-0.41(-0.11)$\\
{}~81.67   & $-0.33(-0.20)$    & ~86.01   & $-0.17(-0.05)$\\
\hline
\end{tabular}

\vspace{.8cm}
\newpage
{\bf Table IV}\\
\vspace{.15cm}

\begin{tabular}{|cc|cc|}
\hline
\multicolumn{4}{|c|}
{$m_0=200$GeV, $m_{1/2}=200$GeV, $A_0=0(200)$GeV}\\
\multicolumn{4}{|c|}{$m_t(M_Z)=150$GeV, $\tan\beta (M_Z)=2.5$}\\
\hline
\multicolumn{4}{|c|}{abs. value of $\mu =311(346)$GeV}\\
\hline
\multicolumn{2}{|c}{$\mu>0$} & \multicolumn{2}{|c|}{$\mu<0$}\\
Tree level mass & Correction (\%) & Tree level mass & Correction (\%)\\
\hline
347.20(377.03)  & $-0.63(-0.92)$    & 326.00(359.74)  & $-0.47(-0.56)$\\
313.99(348.52)  & $-0.42(-0.52)$    & 319.11(352.40)  & $-0.68(-0.85)$\\
142.38(145.77)  & $-0.44(-0.45)$    & 171.71(172.28)  & $-0.18(-0.16)$\\
{}~79.14(~80.46)  & $-0.44(-0.40)$    & ~89.92(~89.80)  & $-0.04(-0.04)$\\
\hline
\end{tabular}

\vspace{.8cm}
{\bf Table V}\\
\vspace{.15cm}

\begin{tabular}{|cc|cc|}
\hline
\multicolumn{4}{|c|}
{$m_0=500$GeV, $m_{1/2}=50$GeV, $A_0=300(700)$GeV}\\
\multicolumn{4}{|c|}{$m_t(M_Z)=180$GeV, $\tan\beta (M_Z)=3$}\\
\hline
\multicolumn{4}{|c|}{abs. value of $\mu =413(419)$GeV}\\
\hline
\multicolumn{2}{|c}{$\mu>0$} & \multicolumn{2}{|c|}{$\mu<0$}\\
Tree level mass & Correction (\%) & Tree level mass & Correction (\%)\\
\hline
429.69(435.28)  & $-2.50(-2.47)$    & 427.14(432.80)  & $-2.16(-2.18)$\\
416.66(422.44)  & $-2.00(-2.00)$    & 417.33(423.09)  & $-2.62(-2.60)$\\
{}~33.75(~33.84)  & $-0.23(-0.22)$    & ~49.88(~49.80)  & $+0.17(+0.17)$\\
{}~16.94(~17.04)  & $-2.39(-2.42)$    & ~23.65(~23.63)  & $+0.19(+0.18)$\\
\hline
\end{tabular}
\end{center}
\hfill
\newpage
{\bf Table Captions}

\vspace{1cm}
{\bf Table I}: Tree level ($m_i$) and radiatively corrected
($m_i^\prime$) neutralino masses (absolute values are shown) for the
inputs values $m_t$, $A_0$, $m_0$, $m_{1/2}$, $\tan \beta$ and $\mu$
shown in the table. The mass corrections \% appear in the last column.

\vspace{1cm}
{\bf Table II}: Tree level neutralino masses (absolute values are shown) and
the resulting radiative corrections \% for $A_0=0$GeV and $A_0=200$GeV
respectively and for values of $m_0$, $m_{1/2}$, $m_t(M_Z)$ and
$\tan\beta$ as shown in the table. The absolute value of $\mu(Q)$
is also given (see main text for details).

\vspace{1cm}
{\bf Table III}: Same as in Table II for the no-scale case: $A_0=0$GeV
and $m_0=0$GeV and for $m_{1/2}$, $m_t(M_Z)$ and $\tan\beta$ as shown in
the table. The numbers in brackets are the mass corrections in the
absence of wave function renormalization effects.

\vspace{1cm}
{\bf Table IV}: Same as in Table II, for different input values for the
parameters $A_0$, $m_0$, $m_{1/2}$, $m_t$ and $\tan\beta$.

\vspace{1cm}
{\bf Table V}: Same as in Table II, for different input values for the
parameters $A_0$, $m_0$, $m_{1/2}$, $m_t$ and $\tan\beta$.

\vspace{2cm}

{\bf Figure Caption}

\vspace{1cm}
{\bf Fig. 1}: The dominant one loop top-stop contribution to the
neutralino two point function.
\end{document}